\documentclass[12pt,aps,amsmath,amssymb, showpacs]{revtex4}
\usepackage{graphicx}
\usepackage{bm}

\newcommand{\beq}{\begin{equation}}
\newcommand{\beqn}{\begin{eqnarray}}
\newcommand{\eeq}{\end{equation}}
\newcommand{\eeqn}{\end{eqnarray}}

\begin{document}

\title{Analysis of Oscillons in the $SU(2)$ Gauged Higgs Model}
\author{Evangelos I. Sfakianakis}
\email{esfaki@mit.edu}

\affiliation{Center for Theoretical Physics  and Department of Physics,
\\ Massachusetts Institute of Technology, Cambridge, MA 02139, USA}
\date{\today}
\begin{abstract}
A stable oscillon was found in the (classical) analysis of the SU(2) gauged Higgs model under a 2:1 ratio between the Higgs and W masses, both analytically and numerically. We extend these results by constructing oscillons for all values of the Higgs mass and examining their stability, finding them to be stable when the Higgs mass is almost equal to twice the W mass. We also investigate the dependence of the oscillon lifetime on the mass of the Higgs, seeing different behavior for different regions of the Higgs mass. Close to the 2:1 mass ratio the interaction of the two fields leads to either a stable oscillon or to its quick decay, based on its amplitude and the exact masses of the fields. Moving further away from this ratio the fields seize to interact and the oscillon disperses.
Furthermore, we show that the oscillon lifetime scales quadratically with the small amplitude parameter which is linked to the small amplitude expansion that we use to construct the oscillon. Finally, we list all families of oscillon solutions that can be formally constructed in this model and discuss their properties. The results can be easily extended to other composite oscillons in particle physics and beyond.

\end{abstract}
\pacs{Preprint MIT-CTP 4412}
\maketitle

\section{Introduction}

The formation of long-lived localized packets of energy is a common feature in nature. The literature of mathematics and physics is full with cases where nonlinearities in field theories lead to stable localized configurations. Relativistic field theories have been known to exhibit such phenomena
We can divide them in time-independent and time-dependent. Their longevity can be due to conserved charges or an effect of the interplay between nonlinearity and dispersion.

We consider a specific type of localized object called an oscillon. Oscillons are localized in space, but oscillate in time  (hence the name) and they do not posses any known conserved charges. In general they are not completely stable, in fact analytical and numerical arguments show an exponentially small radiative tail \cite{tail, radnum, rad1,rad2}. However they are found numerically to have lifetimes that can be orders of magnitude longer than the natural timescales in the theory, such as the oscillation period. This can persist even when they are placed in an expanding background \cite{exp}. If one considers quantum mechanical effects, the radiation of the oscillons is increased \cite{mark}, leading to a shorter lifetime. 

Most documented cases of oscillons involve toy model scalar field theories, where various analytical and numerical techniques have been used to describe their properties, such as their size and lifetime \cite{theory1,theory2,theory3}. Perhaps the most physically interesting characteristic of oscillons is their "natural" emergence in a wide range of scenarios, including inflation \cite{infl1,infl2,infl3}, bubble collisions \cite{bubble}, an expanding system with quasi-thermal initial conditions \cite{emerge} and the interaction of a double well system with a heat bath \cite{bath}.  Finally composite oscillons have been found to  arise in a hybrid inflation setting \cite{composite}.

Recent attempts to find oscillons in the Standard Model lead to the investigation of the electroweak sector. An oscillon in the SU(2) gauged Higgs model was found both numerically \cite{num} and analytically \cite{su2} when the Higgs and W fields have a $2:1$ mass ratio. Also, numerical studies revealed a stable oscillon in the full $SU(2) \times U(1)$ sector \cite{noah1} and \cite{noah2}. Furthermore, simulations have shown that during the electroweak symmetry breaking in an expanding universe, oscillons can emerge and capture a significant amount of the energy of the system \cite{su2cosmo}.

In \cite{su2} Farhi et al construct an oscillon in a restricted version of the $SU(2)$ gauged Higgs model and calculate the oscillon lifetime as a function of the Higgs mass. We explain the lifetime behavior observed by Farhi et al. and thoroughly investigate the whole region of Higgs masses, constructing all possible oscillon solutions, in the small amplitude expansion, and characterizing their stability. Wherever possible we provide intuitive arguments, supported by analytical calculations and numerical simulations. Our analysis can be applied to any case of composite oscillons, meaning that the use of the $SU(2)$ model is in no way restrictive! In fact reducing the analysis of a complicated system, such as the $SU(2)$ model, to simple arguments makes our treatment easily transferable.

Our analysis of of the oscillon's behavior for all values of the Higgs mass reveals regions of qualitatively different behavior.  We formulate the solutions using a small amplitude expansion with parameter $\epsilon \ll 1$. The size of the oscillon is $O(1/\epsilon)$ and the dynamics happens on a timescale of $O(1/\epsilon^2)$. We formally construct new oscillons solutions, who are shown to be unstable. Close to the $2:1$ mass ratio, a formal oscillon solution exists and is found to be stable for a region of Higgs masses, where the Higgs mass is parametrized as $m_H^2 = 4m_W^2 + \epsilon^2 \Delta m_H^2$. Outside the stable region, the two fields that form the oscillon interact and depending on the available decay phase space, the Higgs field transfers its energy into the W or vice versa. As we move further away from the $2:1$ mass ratio, the oscillon is described as a sum of two non-interacting fields, both evolving dominated by dispersion.

The rest of the paper is organized as follows. In Section II and III we present the model and formulate the small amplitude expansion in the various regions of the Higgs mass.The region of masses is divided using the small parameter $\epsilon$, characterizing each region by the deviation of the mass ratio between the Higgs and W masses from 2. In Section IV the oscillon profiles for each oscillon solution are numerically derived. In Sections V and VI, we take a closer look at the evolution of the oscillon, including its lifetime and the interaction between the two constituent fields, for all regions of the Higgs mass and the expansion parameter. Conclusions follow in Section VIII.


\section{Model}

We follow the analysis presented in \cite{num,su2} and will repeat the basic results for completeness.
We consider the Lagrangian of an SU(2) gauge theory coupled to a Higgs doublet in 3+1 dimensions 
\beq
L=\left [  -{1\over 2} Tr~F^{\mu \nu}F_{\mu\nu} + {1\over 2} Tr~(D^\mu \Phi)^\dagger D_\mu \Phi - {\lambda\over 4} (Tr~ \Phi^\dagger \Phi-v^2)^2 \right ] 
\eeq
where $F_{\mu\nu}=\partial_\mu A_\nu-\partial_\nu A_\mu - i ~g [A_\mu,A_\nu]$ , $D_\mu \Phi = (\partial_\mu - igA_\mu)\Phi$ , $A_\mu =A_\mu ^\alpha \sigma^\alpha /2$ , $\Phi(x,t) = \left (
\begin{array} {cc}
\phi_2^* & \phi_1 \\ 
-\phi_1^* & \phi_2 \end{array} {}  \right )$, $\sigma^\alpha$ are the Pauli matrices and
$ds^2=dt^2-dx^2$.
\newline

The theory contains three vector bosons (W's) with mass $m_W=gv/2$ and a Higgs particle with mass $m_H=v \sqrt {2\lambda}$.

In order to be able to search for non-trivial solutions to the equations of motion we will constrain the Lagrangian by using the spherical symmetry \cite{ansatz, ginv}. The spherical Ansatz is given by expressing the gauge field $A_\mu$ and the Higgs field $\Phi$ in terms of six \textbf{real} functions $a_0(r,t)~,~a_1(r,t)~,~\alpha(r,t)~,~\gamma(r,t)~,~\mu(r,t)$ and $\nu(r,t)$
\begin{eqnarray} 
A_0(\vec x,t) &=& {1\over 2g} a_0(r,t) \vec \sigma \cdot \hat x 
\\
 A_i(\vec x,t)& =& {1\over 2g} \left [ a_1(r,t) \sigma \cdot \hat x \hat x_i + {\alpha (r,t) \over r} (\sigma_i -  \vec \sigma \cdot  \hat x \hat x_i )   + {\gamma(r,t)\over r} \epsilon _{i j k} \hat x_j \sigma _k  \right ]
\\
\Phi (\vec x ,t) &=& {1\over g} [ \mu(r,t) + i \nu (r,t) \vec \sigma \cdot \hat x] 
\end{eqnarray}
where $\hat x$ is the unit vector in the radial direction. The conditions at the origin are $a_0,~\alpha,~a_1-\alpha /r,~\gamma /r ,~\nu \rightarrow 0$ as $r\rightarrow 0$. The reduced Lagrangian is
\begin{eqnarray}\nonumber
 L(r,t) =  {4 \pi \over g^2} \left [ -{1\over 4} r^2 f^{\mu\nu} f_{\mu\nu} + (D^\mu \chi ) ^* (D_\mu \chi) +r^2 (D^\mu \phi)^* (D_\mu \phi)  
 \right.
 \\ \left.   - {1\over r^2} (|\chi |^2-1)^2 -{1\over 2}(|\chi |^2+1)|\phi |^2 -Re(i\chi^* \phi^2) -{\lambda\over g^2} r^2 \left ( |\phi |^2-{g^2v^2\over 2} \right ) ^2  \right ]
\end{eqnarray}
where the indices run over 0 and 1, $ds^2=dt^2-dr^2$ and we define $ f_{\mu\nu}=\partial_\mu a_\nu-\partial_\nu a_\mu$ , $\chi = \alpha +i(\gamma -1)$ , $\phi=\mu+i\nu$ , $D_\mu \chi = (\partial_\mu-a_\mu)\chi$ and $D_\mu \phi = (\partial_\mu-{i\over2}a_\mu)\phi$.

This reduced theory has a residual $U(1)$ gauge invariance consisting of gauge transformations of the form $exp(i \Omega (r,t) \vec \sigma \hat x /2)$ with $\Omega (0,t) =0$. Under this, the complex scalar fields $\chi$ and $\phi$ have charges $1$ and $1/2$ respectively, $a_\mu$ is the gauge field, $f_{\mu\nu}$ is the field strength and $D_\mu$ is the covariant derivative.

The equations of motion for the reduced theory are 
\beqn
 \partial^\mu (r^2 f_{\mu\nu}) =i [D_\nu \chi^* \chi - \chi^* D_\nu \chi]+{i\over r^2} r^2  [D_\nu \phi^* \phi - \phi^* D_\nu \phi]  
 \\
 \left [D^2 + {1\over r^2} (|\chi|^2-1) +{1\over 2} |\phi|^2 \right ] \chi = - {i\over 2} \phi^2 
  \\
\left [ D^\mu r^2 D_\mu + {1\over2} (|\chi|^2+1) +{2\lambda \over g^2} r^2 \left ( |\phi|^2-{g^2v^2\over 2} \right ) \right ]\phi = i \chi \phi^* 
\eeqn
The first of these equations is the conserved $U(1)$ current
\beqn
\partial_\mu \left ( i [D_\mu \chi^* \chi - \chi^* D_\mu \chi]+{i\over r^2} r^2  [D_\mu \phi^* \phi - \phi^* D_\mu \phi] \right )=0
\eeqn
A simplification of the spherical ansatz, hereafter referred to as the reduced spherical ansatz, consists of setting the fields $\alpha$, $\nu$ and $a_\mu$ to zero, corresponding to parity-even configurations ($\Phi(-{\bf x})=\Phi({\bf x})$, $A_0(-{\bf x})=A_0({\bf x})$, $A_i(-{\bf x})=-A_i({\bf x})$). This leads to the fields 
\beqn
\phi={gv \over \sqrt 2} \left (1- {\xi \over r} \right )
\\
\chi = i (\eta -1)
\eeqn
where $\xi$ and $\eta$ are real functions of $r$ and $t$. The equations of motion become
\beqn
\partial^\mu \partial_\mu \xi +2\lambda v^2 r \left [ {\xi \over r} -{3\over r} \left ({\xi \over r}\right )^2 +{1\over 2} \left ({\xi \over r}\right )^3 \right ] - {\eta^2\over 2r} \left (1-{\xi \over r} \right )=0
\\
\partial^\mu \partial_\mu \eta +{1\over r^2} (2\eta-3\eta^2+\eta^3)+{g^2v^2\over 4} \left (1-{\xi \over r} \right )^2\eta=0
\eeqn


\section{Small Amplitude Expansion}

We look for oscillons in the weakly nonlinear regime, using the weak-field long-wavelength expansion. The expansion parameter is $\epsilon\ll1$ with $\rho = \epsilon r$ and $\tau = \epsilon^2 t$ as the radial distance and the slow time variable respectively. Keeping the scaling general for now, the fields will be rescaled as $\xi = \delta_\xi u$ and $\eta = \delta_\eta v$. The equations of motion for the two fields become after rescaling
\beqn
\nonumber
{\partial^2 u \over \partial t^2 } + 2 \epsilon ^2 {\partial^2 u \over \partial t \partial \tau } +\epsilon^4 {\partial^2 u \over \partial \tau^2 }+ m_H ^2 u &=& \epsilon ^2 {\partial^2 u \over \partial \rho^2 } + \epsilon \delta_\xi m_H ^2 u^2 3/2\rho + \epsilon {\delta_\eta ^2 \over \delta_\xi} v^2/2\rho
\\
&& - \epsilon^2 \delta_\xi ^2 m_H ^2 u^3 /2\rho^2 - \epsilon^2 \delta_\eta ^2 u v^2 /2\rho^2
\eeqn
\beqn
\nonumber
{\partial^2 v \over \partial t^2 } + 2 \epsilon ^2 {\partial^2 v \over \partial t \partial \tau } +\epsilon^4 {\partial^2 v \over \partial \tau^2 }+ m_W ^2 v &=&
\epsilon^2 \left ( {\partial^2\over \partial \rho^2} - {2\over \rho ^2} \right ) v + \epsilon \delta_\xi 2m_W ^2v u /\rho + \epsilon ^2 \delta_\eta 3 v^2 /\rho^2 
\\
&&- \epsilon ^2 \delta_\eta ^2 v^3 /\rho^2 - \epsilon^2 \delta_\xi m_W ^2 v u^2/\rho^2
\eeqn
For each region of values of the mass ratio between the two fields ($\lambda={m_H^2\over 4m_W^2}$) a different set of scaling parametes for the amplitude of the fields can be chosen to lead to oscillon solutions. We examine each region separately, following and extending the analysis found in \cite{su2}


\subsection{$\boldsymbol{ \lambda \equiv {m_H^2 \over 4 m_W ^2}= 1+O(\epsilon^2)}$}

The chosen field scalings in this case are $\delta_\xi = \delta_\eta = \epsilon$. The physical fields are
\beqn
\phi = {gv \over \sqrt 2} \left [ 1-\epsilon^2 {1 \over \rho  }  Re \{ A(\rho,\tau) e^{-im_Ht} \}  \right ]+O (\epsilon^4)
\\
\chi = -i  \left [ 1-\epsilon Re \{ A(\rho,\tau) e^{-im_Ht} \}  \right ]+O (\epsilon^3)
\eeqn
where the complex amplitudes $A(\rho,\tau)$ and $B(\rho,\tau)$ are defined through the system of coupled equations of the Nonlinear Schroedinger type
\beqn
2im_H {\partial A \over \partial \tau} =-{d^2 A\over d\rho^2}- {1\over 4\rho} B^2
\\
2im_W {\partial B \over \partial \tau} =-{d^2 B\over d\rho^2} + {2\over \rho^2} B  -{m_W^2 \over \rho} AB^* =0
\eeqn
We will call the above set of partial differential equations, envelope equations, since they define the evolution of oscillons on large scales and long times. 
In order to find single frequency oscillatory solutions, we use separation of variables with $ A(\rho,\tau) = a(\rho)e^{im_H \tau/2}$ and $B(\rho,\tau) = b(\rho)e^{im_H \tau/4} $, where $a(\rho)$ and $b(\rho)$ are solutions of the following set of ODE's
\beqn
{d^2 a\over d\rho^2} -m_H ^2 a + {1\over 4\rho} b^2=0
\label{eqn:arho}
\\
{d^2 b\over d\rho^2} -m_W ^2 a - {2\over \rho^2} b  +{m_W^2 \over \rho} ab^* =0
\label{eqn:brho}
\eeqn
We will call these profile equations, since they give the time independent (stationary) solution to the envelope equations, namely the spatial profile of the oscillon. Finally the physical fields are
\beqn
\phi = {gv \over \sqrt 2} \left [ 1-\epsilon^2 {a(\rho) \over \rho  }  cos \left ( m_H t \left ( 1- \epsilon^2/2 \right ) \right ) \right ]+O (\epsilon^4)
\\
\chi = -i  \left [ 1-\epsilon b(\rho) cos \left ( m_W t \left ( 1- \epsilon^2/2 \right ) \right ) \right ]+O (\epsilon^3)
\eeqn
Numerical solutions to the above equations are given in \cite{su2} and the oscillons resulting from these are found to be extremely long-lived.

There are more than one ways to treat the case of $m_H \ne 2m_W$. One is  to take the mass ratio to be very close to $2:1$, that is consider 
\beq
\lambda = {m_H ^2 \over 4m_W^2 }=1+ \epsilon^2 \Delta \lambda= 1+\epsilon^2 {\Delta m_H^2 \over 4 m_W^2}
\eeq
In that case one finds the same form of an oscillon, but the functions $a(\rho)$ and $b(\rho)$ are given by solutions to a slightly different set of equations. In particular the envelope and profile equations for the W field remain unchanged, while the corresponding equations for the Higgs beocme
\beq 
2im_H {\partial A \over \partial \tau} = -{d^2 A \over d\rho ^2} - {1\over 4\rho} B^2 +\Delta m_H^2 A
\eeq
\beq
{d^2 a\over d\rho^2} -(m_H ^2+\Delta m_H^2) a + {1\over 4\rho} b^2=0
\eeq
Before concluding this section, we should note the different asymptotic behavior for a positive or negative value of $\Delta m_H^2$. In the former case the two fields decay as $a \sim e^{-2m_W \rho} /\rho$ and $b\sim e^{-m_W \rho}$, while in the latter case $a\sim e^{-\sqrt {m_H^2 + \Delta m_H^2} \rho} $ and $b\sim e^{-m_W \rho}$, where the in-existence of an exponentially decaying solution for $\Delta m_H^2 \le -m_H^2$ is evident. We will come back to this issue later.

\subsection{$\boldsymbol{m_H-2*m_W=O(1)}$}

At this point we have to choose the two scaling parameters $\delta_\xi$ and $\delta_\eta$. In \cite{su2}, both are taken to be $O(1)$, and the resulting fields are to leading order $u=Re[A(\rho,\tau)e^{-im_H t}]$ and $v=Re[B(\rho,\tau)] e^{-im_W t}]$ where the complex profiles are defined through the PDE's
\beq
2im_H {\partial A \over \partial \tau} = -{\partial ^2 A \over \partial \rho ^2} -{3m_H ^2|A|^2A \over 2\rho^2} + {(6m_W ^2-m_H ^2)|B|^2A \over 2(m_H ^2 - 4m_W ^2) \rho ^2}
\eeq
\beqn
\nonumber
 2im_W {\partial B \over \partial \tau} &=& -{\partial ^2 B \over \partial \rho ^2} - {2B\over \rho^2} + {m_W ^2 (6m_W ^2 - m_H ^2) |A|^2 B \over (m_H ^2 - 4 m_W ^2)\rho ^2} - {(3m_H ^4 - 15m_H ^2 m_W ^4) |B|^2 B \over 4m_H ^2 (m_H^2 -4m_W ^2) \rho ^2}     
\\
&&+ \delta_{m_H,m_W} {m_W^2(m_H^2-2m_Hm_W+2m_W^2)A^2\bar B \over 2m_H (m_H-2m_W)\rho^2}
 \eeqn
where $\delta_{m_H,m_W}=1$ for $m_H=m_W$ and zero otherwise.

The energy of this oscillon is 
\beq
E=\int dr H(r) = {1\over \epsilon} \int d\rho H(\rho) \Rightarrow E \sim {1\over \epsilon}
\eeq
However this does not seem to be the only choice of scaling parameters. If one chooses instead $\delta_\xi = O(1)$ and $\delta_\eta = \epsilon$, one ends up with  $u=Re[A(\rho,\tau)e^{-im_H t}]$ and $v=Re[B(\rho,\tau)] e^{-im_W t}]$ where the complex profiles are defined through a different set of PDE's (envelope equations)
\beq
-2im_H {\partial A\over \partial \tau }= {\partial ^2 A \over \partial \rho ^2} + {14 m_H ^4\over 8 \rho ^2} |A|^2A
\eeq
\beqn
\nonumber
-2im_W {\partial B\over \partial \tau } &=& {\partial ^2 B \over \partial \rho ^2} - {2B\over \rho ^2} + {2m_W ^2m_H ^2 \over (16 m_W ^2 m_H ^2 - m_W ^4)} |A|^2 B
\\
&&+ {6m_W ^2  \over 4\rho ^2} |A|^2 B - {1\over 8 \rho ^2} 3m_H^2 |A|^2A
\eeqn
The interesting fact about this scaling choice, is that it leads to a family of solutions, where the two fields change in a different way as one changes $\epsilon$. Furthermore, the resulting profile equations, when one uses separation of variables in the above set of envelope equations, can be solved sequentially as two de-coupled ODE's. Also, choosing different scaling parameters for the two fields leads to losing the "Lagrangian structure" of the resulting PDE's, that means that they cannot be derived as equations of motion from some effective Lagrangian. Finally, the actual physical fields are very different (since one is suppressed by a factor of $\epsilon$). Because one of the fields is rescaled by an $O(1)$ factor, it will dominate the energy, which will again be $E\sim {1\over \epsilon}$

We'll call this scaling asymmetric, so as to differentiate between the previous scaling, where two fields were scaled by the same number (hence symmetrically).

The study of the oscillon solutions found in this region will be presented in the Appendix.

\subsection{$\boldsymbol{\lambda = 1+O(\epsilon)}$}

Finally in this case, using the same general scaling for the field amplitudes, the equations of motion become
\beqn
\nonumber
{\partial^2 u \over \partial t^2 } &+& 2 \epsilon ^2 {\partial^2 u \over \partial t \partial \tau } +\epsilon^4 {\partial^2 u \over \partial \tau^2 } + m_H ^2 u = \epsilon ^2 {\partial^2 u \over \partial \rho^2 } + \epsilon \delta_\xi m_H ^2 u^2 3/2\rho + \epsilon {\delta_\eta ^2 \over \delta_\xi} v^2/2\rho - \epsilon^2 \delta_\xi ^2 m_H ^2 u^3 /2\rho^2
\\
&& - \epsilon^2 \delta_\eta ^2 u v^2 /2\rho^2  + \epsilon \Delta m_H ^2 u + \epsilon^2 \delta_\xi (3/2\rho) \Delta m_H ^2 u^2 - \epsilon^2 \delta_\xi ^2 (1/2\rho^2) \Delta m_H ^2 u^3
\eeqn
\beqn
\nonumber
{\partial^2 v \over \partial t^2 } + 2 \epsilon ^2 {\partial^2 v \over \partial t \partial \tau } +\epsilon^4 {\partial^2 v \over \partial \tau^2 }+ m_W ^2 v &=&
\epsilon^2 \left ( {\partial^2\over \partial \rho^2} - {2\over \rho ^2} \right ) v + \epsilon \delta_\xi 2m_W ^2v u /\rho 
\\
&&+ \epsilon ^2 \delta_\eta 3 v^2 /\rho^2 - \epsilon ^2 \delta_\eta ^2 v^3 /\rho^2 - \epsilon^2 \delta_\xi m_W ^2 v u^2/\rho^2
\eeqn
Looking at the first equation, we focus our attention on the term $\epsilon \Delta m_H ^2 u $. This means that, whatever the scaling, the $O(\epsilon)$ equation for $u$ will be of the form
\beq
\epsilon u_{1,tt} + m_H ^2 \epsilon u_1 = O(\epsilon)
\eeq
The $O(\epsilon)$ terms are different, based on our choice of scaling parameters $\delta_\xi$ and $\delta_\eta$. All possible such terms are the following:  
\beq
 \epsilon \delta_\xi m_H ^2 u^2 3/2\rho + \epsilon {\delta_\eta ^2 \over \delta_\xi} v^2/2\rho + \epsilon \Delta m_H ^2 u 
 \eeq
To cancel the resonance at this order, the terms on the right-hand side that oscillate with the natural frequency of the ODE ($m_H$) have to sum to zero (important note: $m_H = 2m_W$ to zeroth order). Looking at the three terms, the first does not have a part oscillating with frequency $m_H$, while the other two do. So these two terms must occur at the same order and cancel each other, which leads to the choice $\delta_\eta ^2 = \delta_\xi$.

The $O(\epsilon)$ part of the expansion now reads (keeping only resonant terms)
\beqn
u_{1,tt} + m_H ^2  u_1 &=&  m_H ^2 u_0 ^2 3/2\rho +  v_0 ^2/2\rho + \Delta m_H ^2 u_0  
\\
u_{1,tt} + m_H ^2  u_1 &=& (1/8 \rho) [2|B|^2 + (B^2 e^{-2im_W t} +cc )]+ (\Delta m_H^2 /4) (A e^{-im_H t} + cc)
\eeqn
Canceling the resonant terms gives 
\beq
(1/2 \rho)B^2+(\Delta m_H^2 )A = 0 \Rightarrow A=-{1\over 2 \rho \Delta m_H ^2 } B^2
\eeq
If we only wish to use integer powers of $\epsilon$, the first acceptable solution is to take both to be $O(1)$.

The equations hence become
\beqn
\nonumber
{\partial^2 u \over \partial t^2 } &+& 2 \epsilon ^2 {\partial^2 u \over \partial t \partial \tau } +\epsilon^4 {\partial^2 u \over \partial \tau^2 } + m_H ^2 u = \epsilon ^2 {\partial^2 u \over \partial \rho^2 } + \epsilon  m_H ^2 u^2 3/2\rho + \epsilon v^2/2\rho - \epsilon^2   m_H ^2 u^3 /2\rho^2 
\\
&&- \epsilon^2   u v^2 /2\rho^2  + \epsilon \Delta m_H ^2 u + \epsilon^2  (3/2\rho) \Delta m_H ^2 u^2 - \epsilon^2  (1/2\rho^2) \Delta m_H ^2 u^3
\eeqn
\beqn
\nonumber
{\partial^2 v \over \partial t^2 } + 2 \epsilon ^2 {\partial^2 v \over \partial t \partial \tau } +\epsilon^4 {\partial^2 v \over \partial \tau^2 }+ m_W ^2 v &=&
\epsilon^2 \left ( {\partial^2\over \partial \rho^2} - {2\over \rho ^2} \right ) v + \epsilon 2m_W ^2v u /\rho
\\
&& + \epsilon ^2  3 v^2 /\rho^2 - \epsilon ^2  v^3 /\rho^2 - \epsilon^2 m_W ^2 v u^2/\rho^2
\eeqn
The first order equation for $v$ is 
\beq
v_{1,tt} + m_W ^2  v_1 =    2m_W ^2v_0 u_0 /\rho = (m_W ^2 /2)[ (A \bar B e^{-i(m_H-m_W)t} + cc)+ AB e^ {-i(m_H+m_W)t} +cc)]
\eeq
The first term is the resonant one, and canceling it, means setting it to zero, which means 
\beq
A=B=0
\eeq
This means, that in order to avoid resonances at first order, we cannot choose $\delta_\xi = 1$. The next possible integer pair of rescaling parameters is
$\delta_\xi = \epsilon ^2 $ and $\delta_\eta = \epsilon$. These do not create any terms for $v$ at $O(\epsilon)$. The equations become in this case
\beqn
\nonumber
{\partial^2 u \over \partial t^2 } &+& 2 \epsilon ^2 {\partial^2 u \over \partial t \partial \tau } +\epsilon^4 {\partial^2 u \over \partial \tau^2 }+ m_H ^2 u = \epsilon ^2 {\partial^2 u \over \partial \rho^2 } + \epsilon^3  m_H ^2 u^2 3/2\rho + \epsilon v^2/2\rho - \epsilon^6   m_H ^2 u^3 /2\rho^2 
\\
&&- \epsilon^4   u v^2 /2\rho^2  + \epsilon \Delta m_H ^2 u + \epsilon^4  (3/2\rho) \Delta m_H ^2 u^2 - \epsilon^6  (1/2\rho^2) \Delta m_H ^2 u^3
\eeqn
\beqn
\nonumber
{\partial^2 v \over \partial t^2 } + 2 \epsilon ^2 {\partial^2 v \over \partial t \partial \tau } +\epsilon^4 {\partial^2 v \over \partial \tau^2 }+ m_W ^2 v &=&
\epsilon^2 \left ( {\partial^2\over \partial \rho^2} - {2\over \rho ^2} \right ) v + \epsilon^3 2m_W ^2v u /\rho 
\\
&& + \epsilon ^3  3 v^2 /\rho^2 - \epsilon ^4  v^3 /\rho^2 - \epsilon^4 m_W ^2 v u^2/\rho^2
\eeqn
From the equation for $v$, there are no nonlinear terms at order $\epsilon^2$, meaning that it is impossible to cancel the resonance at this order caused by the term "$2\epsilon^2 v/\rho$". This is true for all higher order scalings, meaning that canceling the first order resonance in $u$ leads to a resonance in $v$ that cannot be canceled. So there is no small amplitude expansion with $\tau=\epsilon^2 t$ and $\rho = \epsilon r$ and fields rescaled as integer powers of $\epsilon$ in this region of the mass ratio.


\section{Numerical Solution of Oscillon Profiles for $\boldsymbol{\lambda=1+O(\epsilon^2)}$}

We will now present the solutions to the profile equations \ref {eqn:arho} and \ref {eqn:brho} for a range of values of the mass mismatch parameter $\Delta m_H ^2$, obeying the boundary conditions $a(\rho) \to 0$  and $b(\rho)/\rho \to 0$ as $\rho \rightarrow 0$ and both decaying exponentially at infinity.

\begin{figure}[] 
   \includegraphics[width=4in]{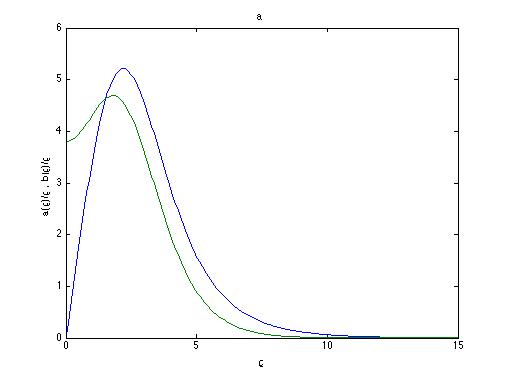} 
   \caption{Profiles of $a(\rho)/\rho$ (green) and $b(\rho)/\rho$ (blue) in the 2:1 mass ratio case. The asymptotic behavior of the two curves at $\rho \to 0$ and $\rho \to \infty$ is evident.}
   \label{fig:0node}
\end{figure}

In \cite{su2} the solutions to the ODE's with $\Delta m_H ^2=0$ were found and fall into a discrete set characterized by the number of nodes in $b(\rho)$. We will focus on the "fundamental" solution with zero nodes. Fig. \ref{fig:0node}  shows the field profiles in the case of the 2:1 mass ratio and it is identical to the ones given in \cite{su2}.

\begin{figure}
\centering
\begin{tabular}{cc}
\includegraphics[width=3.8in]{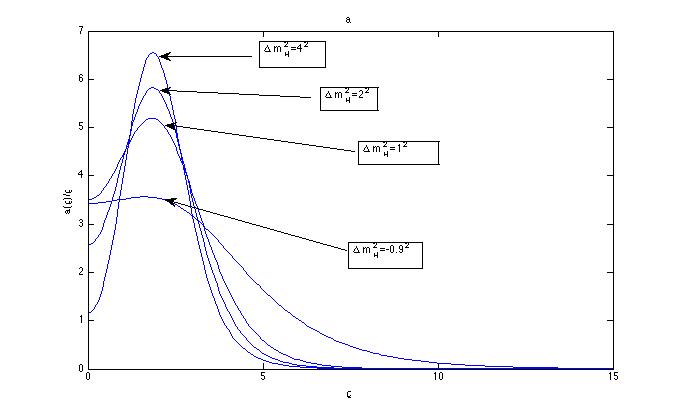}&
\includegraphics[width=3.1in]{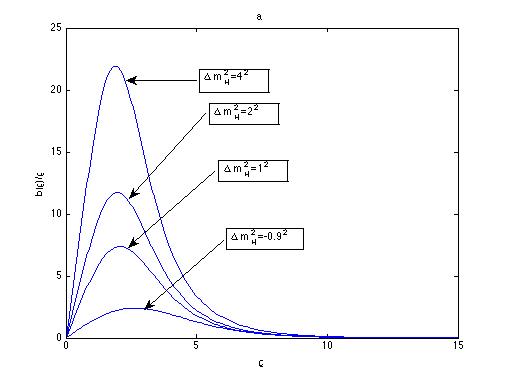} \\
\end{tabular}
   \caption{Profiles of $a(\rho)/\rho$ (left) and $b(\rho)/\rho$ (right) for different values of the mass ratio. The profiles become lower in amplitude and larger in width as the mass mismatch gets smaller, until they dissolve for $\Delta m_H ^2 \to -1$. The change is more dramatic in $b(\rho)/\rho$.} 
   \label{fig:a0node}
\end{figure}

\begin{figure}

\begin{tabular}{cc}
\includegraphics[width=3.5in]{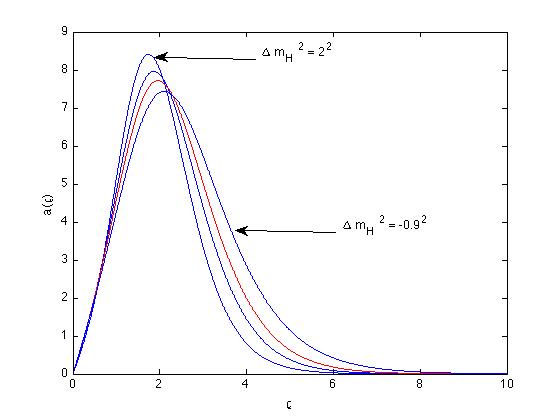}&
\includegraphics[width=3.5in]{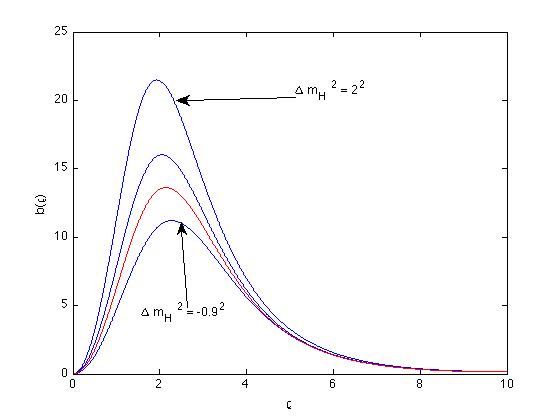} \\
\end{tabular}
   \caption{Profiles of $a(\rho)$ (left) and $b(\rho)$ (right)   for different values of the mass ratio. The profiles become lower in amplitude and larger in width as the mass mismatch gets smaller, until they dissolve for $\Delta m_H ^2 \to -1$. The change is more dramatic in $b(\rho)$.} 
   \label{fig:ar0node}
\end{figure}

We now start to change the mass ratio, by adjusting  $\Delta m_H ^2$. It is worth saying that the actual mass ratio depends on $\Delta m_H ^2$ as well as the expansion parameter $\epsilon$.
There are two qualitatively different regimes of $\Delta m_H ^2$. On the positive side, there is no a priori upper limit for this number. On the negative side, the behavior of the profile equations will change at $\Delta m_H ^2=-4 m_W^2 \Rightarrow \Delta \lambda =-1$, where the linear term in the profile equation for $a(\rho)$ changes sign, and the asymptotic behavior at infinity goes from being exponentially suppressed to oscillatory.  The existence of this transition in the behavior of the profile equations can be explained by the nature of the small amplitude expansion. Let's consider the frequency of the Higgs field, both inside the oscillon and in its free state. The squared natural frequency of the Higgs field is $\lambda 4 m_W^2= 4m_W^2 (1+\Delta \lambda)$, while the squared frequency of the Higgs part of the oscillon is $4m_W^2 (1-\epsilon^2)$, which for $\Delta \lambda< -1$ becomes larger than the frequency of the free field. An intuitive necessary condition for the existence of oscillons, regardless of their structure or dimensionality, is the requirement that the frequency of the oscillon be lower than the natural frequency (or mass in particle terms) of the free field. Thinking of the time dependent part of the field as a ball rolling inside a potential well this condition is translated into the requirement that the potential gets shallower than quadratic at some point. As long as the field can probe this point, oscillons can exist, although their stability requires further analysis. It thus follows that if $\Delta \lambda < -1$ oscillons cannot exist, which is automatically captured by our envelope equations, since they do not admit localized solutions in that regime.

We explored the space $-0.9^2<\Delta m_H ^2<4^2$, where we were able to find well behaved solutions, at least for the fundamental zero-mode solution. The behavior of profiles of the two fields are shown in Fig. \ref{fig:a0node} and \ref{fig:ar0node}. Qualitatively, the profiles become larger as the mass mismatch parameter gets larger. Furthermore $b(\rho)$ is more affected by the change in the mass than $a(\rho)$, although the ODE for $a(\rho)$ is the one that explicitly depends on $\Delta m_H ^2$.
The results seem to indicate that there is no obvious upper cutoff on the positive side of $\Delta m_H ^2$ and that we can approach arbitrarily close to the negative cutoff $\Delta m_H ^2=-4m_W^2$. The profiles for both fields become broader and their amplitude decreases as we lower the mass mismatch parameter $\Delta m_H ^2$. That was expected, since the profiles seize existing at $\Delta\lambda=-1$, so we are seeing a smooth transition from localization to delocalization. Since in our numerical code we require the fields to decay exponentially at infinity, delocalization is manifested in the form of a vanishing amplitude.

\begin{figure}[] 
   \includegraphics[width=4in]{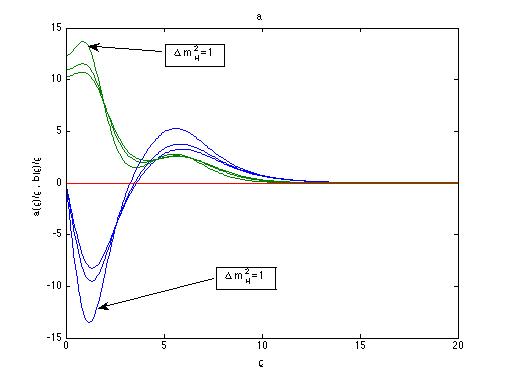} 
   \caption{1-node profiles of $a(\rho)/\rho$ and $b(\rho/\rho)$ for $\Delta m_H ^2=-0.5^2, 0, 1$. The apparent relative sign between the profiles of the two fields is of no physical importance, since the Higgs oscillates with twice the W frequency, hence the relative sign of the amplitudes changes twice every W period. } 
   \label{fig:1node}
\end{figure}

In order to see whether or not the behavior of the solutions for changing $\Delta m_H ^2$ is accidental for the fundamental solution or generic, we solve the profile equations for the single-node case.
Although we didn't explore the same region of $\Delta m_H ^2$, the existence of solutions and their qualitative behavior is the same as before, as is shown in Fig. \ref{fig:1node}. In this case, the solutions continue to grow larger for increasing $\Delta m_H ^2$ and $b(\rho)$ continues to be affected more by the value of the mass mismatch parameter than $a(\rho)$.

As we move along the $\lambda$ line for a fixed $\epsilon$ the initial energy of the oscillon changes, so we need to calculate the energy of the oscillon, based on the $\epsilon$-expansion. The energy for the reduced set of equations is 
\beqn
\nonumber
E=\int_0 ^\infty    dr H ={4\pi \over g^2}\int_0 ^\infty    dr \left [   \left ( {d\eta \over dt} \right ) ^2 + \left ( {d\eta \over dr} \right ) ^2  + {g^2 v^2\over 2} \left (   \left ( {d\xi \over dt} \right ) ^2  + \left ( {d\xi \over dr} \right ) ^2  \right ) \right .
\\
\left .+ {1\over 2r^2}(\eta^2-2\eta)^2+{g^2 v^2 \over 4}\eta^2\left (1-2{\xi\over r} +{\xi^2 \over r^2} \right ) + \lambda {g^2v^4 \over 4} \xi^2 \left ( 2-{\xi \over r}\right )^2 \right ]
\eeqn
Only the last term depends explicitly on the mass mismatch $\lambda$. Furthermore each term has a different dependence on $\epsilon$. We integrate each terms separately (based on the numerical solution obtained for the $2:1$ case) and keep the $\epsilon$ dependence of each terms explicit. The reason for using the $2:1$ profile is that it makes comparison of our results with \cite{su2} easier.

Before that, let's see the terms that are dominant for small $\epsilon$
\beq
E ={4\pi \over g^2} \left [   \left ( {d\eta \over dt} \right ) ^2   + {g^2 v^2\over 2} \left ( {d\xi \over dt} \right ) ^2   +{g^2 v^2 \over 4}\eta^2 + \lambda {g^2v^2 \over 4} \xi^2 4 \right ]
\eeq
Taking each field to oscillate with its natural frequency (mass), which is correct to lowest order in $\epsilon$ and integrating the Hamiltonian to find the energy
\beq
E= \int_0 ^\infty dr H = \epsilon \int_0 ^\infty d\rho {4\pi \over g^2} \left [ { v^2 g^2 \over 4}\eta_{res} ^2 + \lambda  v^4 g^2 \xi_{res}^2 \right ]
\eeq
where $\eta=\epsilon~ \eta_{res}(\rho) cos(\omega_H t) $ and $\xi=\epsilon ~\xi_{res}(\rho) cos(\omega_W t) $

Evaluating the two integrals numerically and neglecting a constant factor multiplying the whole energy, but concentrating on the relative size of the two terms, we get
\beq
E \sim \epsilon (1+1.15\lambda)
\eeq
From this it is clear how the energy changes as a function of the mass mismatch. Since we define the W mass to be constant, the mass mismatch only affects the mass of the Higgs, so the above term shows also the distribution of the oscillon energy among the two fields. It is important to remember that the interaction energy is higher order in $\epsilon$, and hence does not appear in this formula. By calculating the energy of the oscillon and comparing with this formula, the relative error is $O(\epsilon^2)$ as expected.


\section{Oscillon lifetime}

It is now important to evaluate the lifetime of the oscillon for the different mass mismatch (relative to the 2:1 ratio). In accordance with \cite{su2}, we will evolve the solution within the reduced spherical Ansatz to show stability. If not perturbed, the solution will remain within this Ansatz, so this simplification does not affect the results. Obviously the solution could be unstable if perturbed outside the reduced spherical Ansatz, but this does not invalidate the analysis that only considers this reduced case. In order to check stability in the full spherical Ansatz, the small-amplitude expansion will have to be done for the full set of fields.

The equations that we will be numerically evolving are
\beq
\partial^\mu \partial_\mu \xi + 2 \lambda v^2 r \left [ {\xi \over r} - {3\over 2} \left ( {\xi \over r} \right ) ^2 + {1\over 2}\left ( {\xi \over r} \right ) ^3 \right ]-{\eta^2\over 2r} \left ( 1-{\xi \over r} \right )=0
\eeq
\beq
\partial^\mu \partial_\mu \eta + {1\over r^2} (2\eta -3\eta^2 +\eta^3) + {g^2 v^2 \over 4} \left ( 1-{\xi \over r} \right )^2 \eta=0
\eeq
The numerical method used is discretizing the above equations using a second order finite difference scheme and evolving them explicitly forward in time. We use a finite spatial box having a length of several times the oscillon width with a reflecting boundary at infinity. We have seen that the total energy inside the box is conserved and no numerical instabilities seem to occur. 

Without loss of generality, we choose units such that  $g=\sqrt 2$ and $v=1$, the same as in \cite{su2}. The lifetimes will be calculated using the $\lambda=1$ profile as initial conditions even in the cases of $\lambda \ne 1$, to be consistent with \cite{su2}, since explaining the form of the lifetime plot observed by Farhi et al is among the scopes of the present paper.

\begin{figure}[] 
   \includegraphics[width=5in]{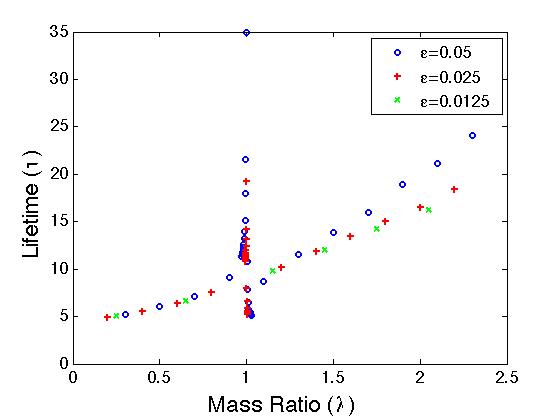} 
   \caption{Lifetime of the oscillon as a function of the mass ratio $\lambda$ for $\epsilon=0.0125$, $\epsilon=0.025$ and $\epsilon=0.05$. Lifetime is measure in units of $\tau = \epsilon^2 t$ and the mass ratio is defined as $\lambda = {m_H \over 2m_W}$ We can see that all curves behave identically, with a small deviation of the $\epsilon=0.05$ curve for large mass ratio. The region around $m_H = 2m_W$ is not very clear in this graph and will be analyzed later.} 
   \label{fig:LifetimeF}
\end{figure}

The lifetime is defined as the time needed for $50\%$ of the initial energy to leave the oscillon region (the oscillon region is defined as the region containing $99\%$ of the oscillon energy initially). We choose three values of the small amplitude parameter, namely $\epsilon=0.0125, ~0.025,~0.05$ and the lifetime plot is shown in Fig. \ref{fig:LifetimeF}. It is important to note, that the vertical axis is plotted in units of the rescaled "slow" time, $\tau = \epsilon^2 t$. This graph is identical to the one presented in \cite{su2} for the $\epsilon = 0.05$ case.

The fact that the dynamics of the oscillon within the small amplitude expansion happens on the slow timescale $\tau$ similarly for each $\epsilon$ is evident from the lifetime plot. This is a general feature of the small amplitude analysis used here, as will be explained later, for all values of the mass ratio. 
Furthermore, lifetimes calculated for three values of $\epsilon$ fall almost on the same curve, except for a small region of values close to $\lambda =1$, which will be analyzed later. There one needs to rescale the horizontal axis for each value of  $\epsilon$ to recover this feature.

Also we see a deviation of the $\epsilon=0.05$ curve from the other two for large mass ratio. By looking closely into the localized field energy as a function of time for this region, we see that the change in lifetime occurs because of a slight change in the decay rate of the Higgs. This explains why the discrepancy happens for large mass ratio, where the oscillon contains more energy in the Higgs field. In this region a single field Higgs oscillon exists (examined in more detail in the Appendix), whose amplitude is $\epsilon$ - independent and falls between the amplitude of the "regular" two field oscillon for $\epsilon=0.025$ and $\epsilon = 0.05$.  This means that the nonlinearity there can be strong enough to affect the decay of the $\epsilon = 0.05$ oscillon, but not the other two ($\epsilon = 0.025$ and $\epsilon = 0.0125$). We will not study this slight discrepancy any further.

Perhaps the most striking feature of the lifetime plot is the dip, occurring at a different point for each $\epsilon$: for $\epsilon = 0.05$ at $\lambda \approx 1.025$ and for $\epsilon=0.025$ at $\lambda \approx 1.006$. Both values of the position of the dip are consistent with $\lambda \approx 1+10\epsilon^2$. This value is taken as an assumption for now and will be explained later.

\begin{figure}
\includegraphics[width=5in]{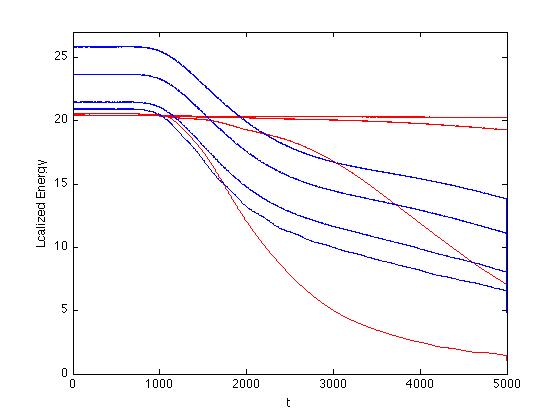}
   \caption{Localized energy as a function of time for $\epsilon = 0.05$ and a heavy Higgs. From top to bottom: $\lambda=1.5, 1.3, 1.1, 1.05$ (blue) and  $\lambda=1.001, 1.002, 1.005, 1.015$ (red). Oscillons with a large mass ratio lose energy at the same rate and larger mass ratio oscillons have a larger initial energy. For oscillons close to the stability point ($m_H=2m_W$) the energy loss ratio goes down as one moves towards the $2:1$ mass ratio. } 
   \label{fig: en1}
\end{figure}

Apart from the lifetime, it is interesting to see how the energy within the initial oscillon region changes with time. We will start with the $\lambda>1$ region, since it includes the counter-intuitive feature that requires an explanation, namely the dip. We can see two different regions in the behavior of the energy evolution, which also occur on the two sides of the dip. In Fig. \ref {fig: en1} we see two qualitatively different regimes, each of them also occurring on either side of the dip. For $\lambda \gtrsim 1+10\epsilon^2=1.025$ the decay curves are almost the same, but with a shifted origin.

For values of $1<\lambda \lesssim 1+10\epsilon^2$, the decay curves are qualitatively different from each other, having a greater decay rate as $\lambda$ increases. On the other hand, when $\lambda$ leaves this region (which coincides with the location of the dip), the behavior changes. The initial time of constant localized energy is the same, and so is (almost) the rate of energy loss. The only difference is the fact that oscillons with different value of $\lambda$ and the same profile have different energies. So it simply takes more time to radiate more energy with the same rate, that is why oscillons at larger $\lambda$ seem to live longer. This is the observational explanation of the dip, the actual physical reason for this behavior will be analyzed in the next section.

\begin{figure}[] 
   \includegraphics[width=5in]{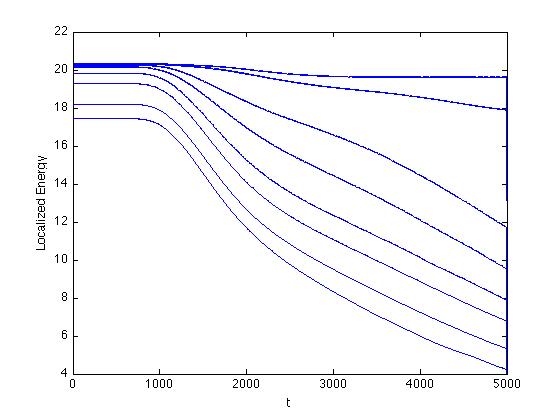} 
   \caption{Localized energy as a function of time  for $\epsilon = 0.05$ and a heavy Higgs. From top to bottom $\lambda = 0.998,~0.997,~0.99,~0.98,~0.95,~0.9,~0.8,~0.7$. Again the decay rate is growing as one moves away from the stability point and freezes, once we exit the $\lambda=1+(\epsilon^2) region$. From that point onward the decay curves are shifted copies of each other.} 
   \label{fig:energy_less_1}
\end{figure}

Looking at the region of $\lambda<1$, the localized energy as a function of time is shown in Fig. \ref{fig:energy_less_1}. For $\lambda \lesssim 1-10\epsilon^2=0.975$ the decay curves are identical, except from being shifted by the different value of $\lambda$. The plateau of the localized energy and the energy loss rate are identical for all mass mismatches. As we move into the $\lambda=1-O(\epsilon^2)$ region, the plateau remains the same. However the energy loss rate reduces as we approach the stability point $\lambda=1$. The oscillon for $\lambda = 0.998 = 1-0.8*\epsilon^2$ shows only an initial loss of energy. This is due to the fact that we start with the oscillon profile of the $\lambda=1$ case, which has more energy than the actual oscillon for $\lambda = 0.998$. What we see is the initial lump shedding its excess energy and relaxing to a stable oscillon configuration.

\begin{figure}[] 
   \includegraphics[width=5in]{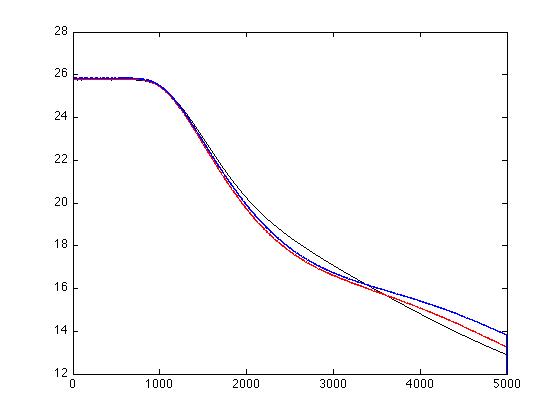} 
   \caption{Comparison between decay curves for $\lambda=0.8$ (black) and $\lambda=1.3$ (red) and ,$\lambda=1.5$ (blue) shifted to a common origin. We see that once shifted the three curves are almost identical. All three belong to the $\lambda=1+O(1)$ region and their similarity indicates a common decay mechanism that is independent of the Higgs mass.} 
   \label{fig:out_comp}
\end{figure}

In Fig. \ref{fig:out_comp} we compare the decay curves of two oscillons outside the $\lambda=1+O(\epsilon^2)$ region in both sides of the stability point. We see that, once shifted, they practically fall on top of each other!

\section{Understanding the different regions of the $\boldsymbol{\epsilon -\lambda}$ plane}

Now that we have a better understanding of the lifetime plot presented initially in \cite{su2}, we can analyze the whole parameter space for different oscillon behavior. We will use numerical simulations both to guide our intuition and to verify our analytical arguments.

\subsection{Large $\boldsymbol{\epsilon}$}

\begin{figure}[] 
   \includegraphics[width=4in]{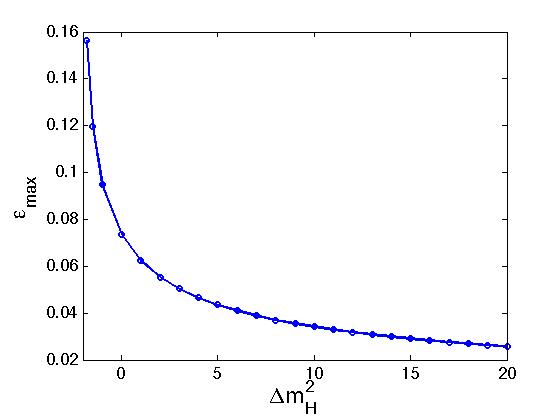} 
   \caption{Maximum $\epsilon$ for the expansion to be valid, as a function of $\Delta m_H ^2$. Oscillons that start above this curve will have a large enough amplitude to probe both minima of the double well Higgs potential.} 
   \label{fig:emax}
\end{figure}

The analysis of the oscillons is based on the fact that the field oscillates back and forth in an anharmonic potential well. However, for large enough value of the field, it will cross over to the second part of the double-well Higgs potential. Between the two fields, the W (in our small amplitude analysis, the $\chi$) will cross over first. The condition for the field to be confined within one well is 
\beq
\epsilon_{\max}  = {1\over max[b(\rho)]}
\eeq
For the oscillon at the $2:1$ mass ratio, the maximum $b(\rho)$ is 13.63, so the maximum $\epsilon$ is  0.073, which is larger (but not by much) to the value 0.05 that was initially used in our simulations as well as in \cite{su2}. We have seen that $b(\rho)$ grows with $\Delta m_H ^2$, so $\epsilon_{max}$ will decrease, as shown is Fig. \ref{fig:emax}.


\subsection{$\boldsymbol{|\lambda-1| >  O(\epsilon ^2)}$ region}

We have seen that lifetime plots include a dip at around $\lambda = 1+10\epsilon^2$. Moreover, beyond that point the lifetime plots have a characteristic behavior. There is a plateau until $t\approx 800$ or $\tau \approx 2$.
Most importantly the decay of the localized energy $dE_{loc} \over dt$ is almost identical in all cases, for values of $\lambda$ on both sides of the stability point $\lambda=1$.

Let's think of the two fields that comprise the oscillon, in the spirit of Feynman diagrams. Since we have restricted the fields to be in the reduced spherical Ansatz, their interaction has been modified, so we can't easily do actual calculations, but we can use the full $SU(2)$ Feynman diagrams to do an order of magnitude estimation.

By examining a Higgs in its rest frame, assuming it has more than twice the mass of the W, the energy-momentum conservation gives us
\beq
m_H=2\sqrt{m_W^2+k^2} \Rightarrow ({k\over 2m_W} )^2={m_H^2 \over 4m_W^2}-1=\lambda -1 
\eeq
where $k$ is the wavenumber of the outgoing W's and $c=1$.

 Quantum mechanically the above calculation will immediately give us the lifetime of the Higgs.
However following the reasoning of \cite{mark}, we realize that in classical field theory, a decaying Higgs field will excite the appropriate wavenumber of the W-field. Hence, these wavenumbers have to be present in the original configuration.

In the small amplitude analysis, the oscillon has a width of $O(1/\epsilon)$, which means that its Fourier transform has a width of $O(\epsilon)$. By assuming there are no wavenumbers beyond some $O(\epsilon)$ value (actually they are there, but are exponentially suppressed), we get 
\beq
{O(\epsilon^2) \over 4m_W^2} = \lambda-1 \Rightarrow \lambda =1+O(\epsilon^2)
\eeq
So, in order to get an effective pumping of the W-field, we need to be in the $\lambda =1+O(\epsilon^2)$ region. Outside this, even a super-heavy Higgs will classically not decay into W's. Even more, the more massive the Higgs, the more exponentially suppressed its decay will be!

This does not apply only to the heavy-Higgs case. Turning the argument around, one can construct a Feynman diagramm including a 4-W vertex, in addition to the Higgs-W interaction. The Feynman graph will include  three ingoing W's and an outgoing Higgs and an outgoing W. Hence, classically we will see a pumping of the Higgs field with a parallel demise of the W field. If we take the initial particles to be at rest, energy-momentum conservation can be written as
\beqn
\nonumber
9m_W^2 &=& m_H^2+k^2+m_W^2+k^2+2\sqrt{(m_H^2+k^2)(m_W^2+k^2)} \Rightarrow
\\
\lambda& =& {m_H^2\over 4m_W^2} = -{3\over 2} \left ( {k\over 2 m_W} \right ) ^2 + {5\over 8} + \sqrt{   -{3\over 4} \left ( {k\over 2 m_W} \right ) ^4 + {3\over 8}  \left ( {k\over 2 m_W} \right ) ^2  +\left ( {3\over 8} \right )^2}
\eeqn
For the long wavelength modes that are present in the oscillon we can Taylor expand the above solution for $k\ll m_W \ll 1$, which gives
\beq
\lambda = 1- \left ( {k\over 2 m_W} \right ) ^2 = 1- O(\epsilon^2)
\eeq
As before, the field interaction is restricted to the region of masses with  $\lambda =1+ O(\epsilon ^2)$. Since the two fields do not interact for $|\lambda-1| >  O(\epsilon ^2)$, each field must decay on its own, due to dispersion. For a gaussian wavepacket in a massive field theory we can write down a general expression for the evolution of the width $\sigma(t)$ that is 
\beq
\sigma (t) ^2-\sigma(0)^2 \sim{ t^2 \over m \sigma(0)^2}
\eeq
For a general wavepacket, the fact that the dispersion will be inversely proportional to the mass and the square of the initial width will persist. In our case this means that 
\begin {itemize}
\item The W field will decay roughly twice as fast as the Higgs
\item Both will decay on a timescale of $1\over \epsilon^2$ which is again the slow timescale $\tau$ of our small amplitude analysis
\end{itemize}

Both features can be seen numerically.

By setting one of the two fields initially to zero and running the simulation, the resulting energy-time plots are identical to when we start with both fields, meaning that the two are practically non-interacting. Furthermore, we can see the W decaying almost twice as fast as the Higgs. The fact that the W decays first and that the mass of the W is fixed, means that initially all decay curves will be identical! This explains the observed behavior, that when shifted, decay curves for different vales of $\lambda$ fall on top of each other.

\begin{figure}
\centering
\begin{tabular}{cc}
\includegraphics[width=3.5in]{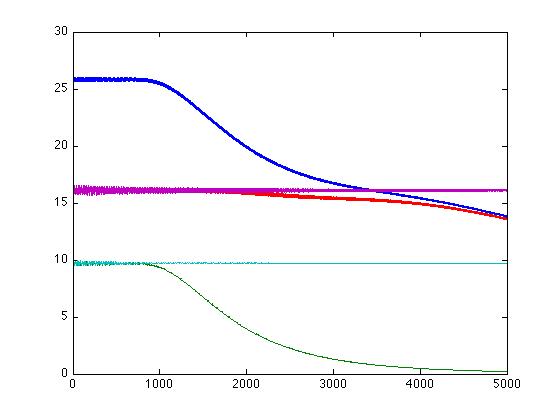}&
\includegraphics[width=3.5in]{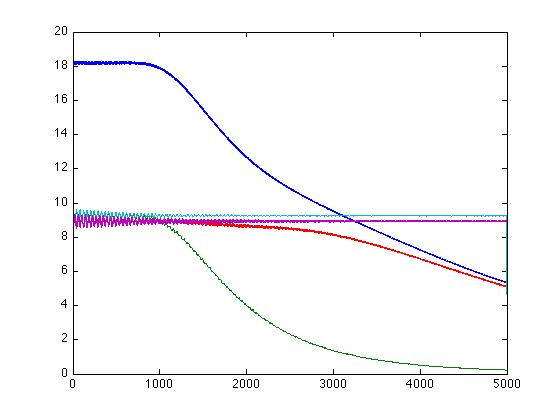} 
\end{tabular}
   \caption{Evolution curves for $\lambda=1.5$ (left) and $\lambda=0.8$ (right). We plot the total localized energy (blue), the localized Higgs energy (red), the total Higgs energy (purple), the localized W energy (green) and the total W energy (cyan). The W decays first, roughly twice as fast as the Higgs, and the total energy shed is initially due to the W decay. Furthermore the total energy in the two fields is separately conserved, pointing towards an extremely suppressed interaction. }
   \label{fig: twofield}
\end{figure}

Simply put, the energy of the oscillon changes with $\lambda$ as $E \sim 1+ 1.15 \lambda$ where the term proportional to $\lambda$ represents the energy in the Higgs field. The W, which has a constant energy content for all values of $\lambda$ decays twice as fast as the Higgs, hence dominates the decay rate initially. By changing $\lambda$ we put more or less energy in the field that decays slowly, meaning that we change the total oscillon energy but not the initial decay rate, so we change the lifetime.

\subsection{$\boldsymbol{\lambda = 1+O(\epsilon ^2)}$ region}

This is the most interesting region, since it is close to the stability point $\lambda=1$ and furthermore there is a well defined small amplitude expansion. 

\begin{figure}
\centering
\begin{tabular}{cc}
\includegraphics[width=3.5in]{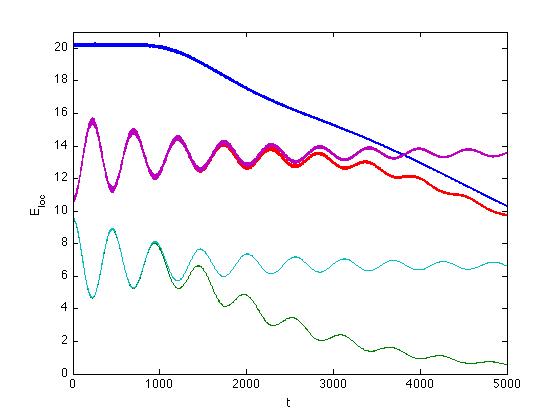}&
\includegraphics[width=3.5in]{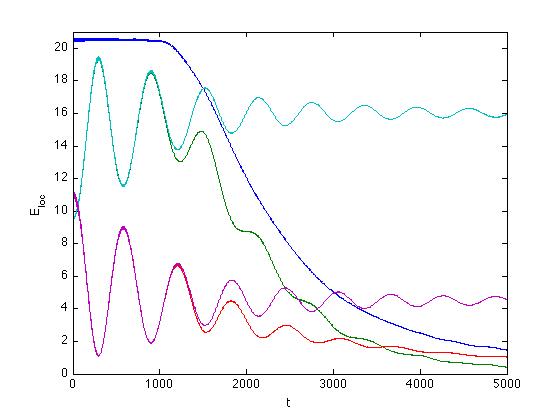} 
\end{tabular}
   \caption{Evolution curves for $\lambda=0.985$ (left) and $\lambda=1.015$ (right) for $\epsilon=0.05$.  We plot the total localized energy (blue), the localized Higgs energy (red), the total Higgs energy (purple), the localized W energy (green) and the total W energy (cyan). We see a continuous transfer of energy between the two fields. A light Higgs ($m_H<2m_W$) is pumped by the W, while a heavy Higgs ($m_H>2m_W$) pumps the W. The latter process is also far more efficient, leading to a quicker overall decay of the oscillon.}
   \label{fig:asym}
\end{figure}

Before going into the stability analysis, we can understand the asymmetry around the $\lambda=1$ point, where we see that  in the $\lambda>1$ region the lifetime falls much faster than in the $\lambda<1$ region in Fig. \ref{fig:asym}. We see two factors that both lead to the very faster demise of the oscillon in the heavy Higgs region. For a heavy Higgs (by "heavy Higgs" we mean $m_H>2m_W$), the Higgs field decays and pumps the W fields very effectively and then the decay of the oscillon happens through dispersion of the W field. In the opposite region ($m_H<2m_W$) the Higgs cannot pump the W, but instead the W pumps the Higgs, which is a process that happens far less effectively. Then the oscillon decay curve follows the dispersion of the Higgs, which is almost twice as heavy as the W, hence disperses with half the rate.

To lowest order in $\epsilon$ the fields in this region are $\xi=\epsilon Re[A(\rho,\tau)e^{-im_Ht}]$ and $\eta=\epsilon Re[V(\rho,\tau)e^{-im_Wt}]$ where the profiles $A$ and $B$ are defined through their envelope equations. According to the small amplitude expansion, processes that occur over a distance scale of $O({1\over \epsilon})$ and over a time scale of $O({1\over \epsilon^2})$ are captured by the envelope equations. Since we expect any instability to involve small wavenumbers (that is long wavelengths) and our simulations give us a lifetime of $O({1\over \epsilon^2})$, we will use the envelope equation to study the behavior of the oscillon in the $\lambda = 1+O(\epsilon ^2)$ region and compare with results from full wave simulations for certain values of the expansion parameter $\epsilon$

Three important observations justify that the envelope equations (that describe the behavior of the oscillon over long time-scales and long wavelengths) is enough to characterize the instability around the $\lambda=1$ point.

\begin{figure}
\centering
\begin{tabular}{cc}
\includegraphics[width=3.3in]{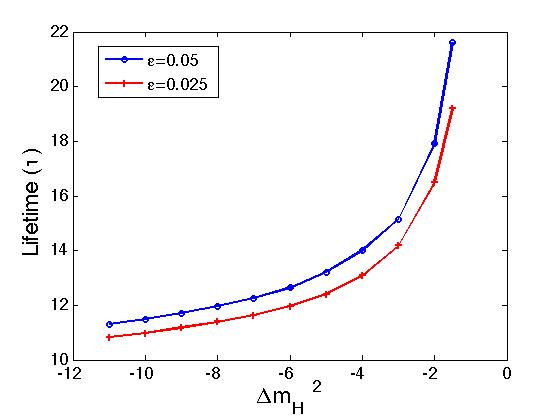}&
\includegraphics[width=3.3in]{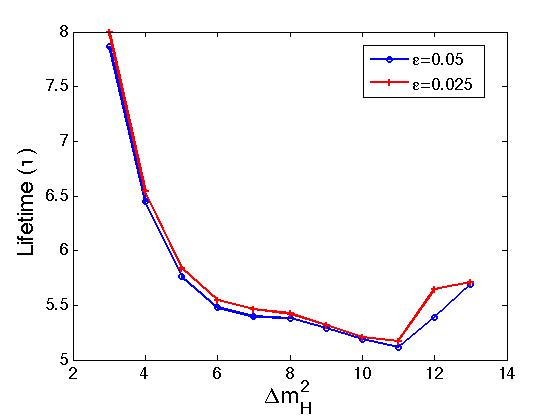} 
\end{tabular}
   \caption{Lifetime (in $\tau$) of the oscillon around the stability point as a function of the mass mismatch $(\Delta m_H ^2 / 4m_W^2)$ for $\epsilon=0.025$ and $\epsilon=0.05$. Lifetime is measured in units of $\tau = \epsilon^2 t$ and the mass mismatch parameter is $\Delta m_H^2$, leading to $m_H^2 = 4m_W ^2 + \epsilon^2 \Delta m_H^2$. We can see that once rescaled the curves follow each other very closely, since both follow the same set of envelope equations.}
   \label{fig:lifetime_nls}
\end{figure}

\begin {itemize} 
\item The decay for two choices of $\epsilon$, namely $\epsilon=0.025$ and $\epsilon=0.05$ occurs over timescales of $O(\epsilon^2)$, which is exactly the scaling of time in the small-amplitude expansion
\item Thinking about the decay of an oscillon in terms of Feynman graphs, we conclude that the important processes in this case involve wavelengths of $O(\epsilon)$, which is exactly the scaling of space in the small-amplitude expansion.
\item Zooming in on the $\lambda = 1+O(\epsilon^2)$ region of Fig.\ref{fig:LifetimeF} and scaling out the $\epsilon$ dependence from the $\lambda$ axis, the lifetime curves for both $\epsilon$ values are similar, as seen in Fig. \ref{fig:lifetime_nls}
\end {itemize}

The envelope equations in the region $\lambda = 1+O(\epsilon^2) = 1+\epsilon^2 \Delta \lambda = 1+ \epsilon^2 {\Delta m_H ^2 \over 4 m_W ^2}$ 
\beqn
2im_H {\partial A \over \partial \tau} = -{\partial ^2 A\over \partial \rho ^2} - {1\over 4\rho} B^2 + \Delta m_H ^2 A
\\
2im_W {\partial B \over \partial \tau} = -{\partial ^2 B\over \partial \rho ^2} + {2\over \rho^2} B - {m_W ^2 \over \rho } A\bar B
\eeqn
We make the substitution $A \to A e^{i\Omega \tau}$ and  $B \to B e^{i\Omega \tau /2} $ and the envelope equations become 
\beqn
2im_H {\partial A \over \partial \tau} = -{\partial ^2 A\over \partial \rho ^2} - {1\over 4\rho} B^2 + \Delta m_H ^2 A + 2m_H \Omega A
\\
2im_W {\partial B \over \partial \tau} = -{\partial ^2 B\over \partial \rho ^2} + {2\over \rho^2} B - {m_W ^2 \over \rho } A\bar B + m_W \Omega B 
\eeqn
The stationary solutions to these equations ($\partial / \partial \tau =0$) are the oscillon amplitudes $a(\rho,\Omega)$ and $b(\rho,\Omega)$. Evem though it seems that we introduced one extra free parameter, in effect constructing an extra family of oscillons, we can always rescale the fields and the space time variables by a constant factor, in order to set $\Omega = m_W$. In order to apply the stability results as they are found in the literature, we will keep $\Omega$ explicit for now.

Following the reasoning found in  \cite{stab} and \cite{stab2}, which is an extension of the Vakhitov-Kolokolov criterion for soliton stability \cite{VK} and \cite {VK2} , oscillons are stable with respect to perturbations (solutions to the above equation are non-growing) if 
\beq
{\partial Q_S \over \partial \Omega_S} >0
\eeq
where $Q_S$ is the particle number invariant 
\beq
 Q = \int_0 ^\infty d\rho (|A|^2 + {1\over 4}|B|^2) 
 \eeq
calculated at the stationary solution $(A_s=a,B_s=b)$. Although this result is obtained for a system of PDE's that differs from our own in the fact that ours has space-dependent coefficients, we will use the criterion as described above and test its predictions by a full wave simulation. We are currently working towards formally extending the validity of the stability analysis for a more general class of systems, like the one presented here.

\begin{figure}[] 
   \includegraphics[width=4in]{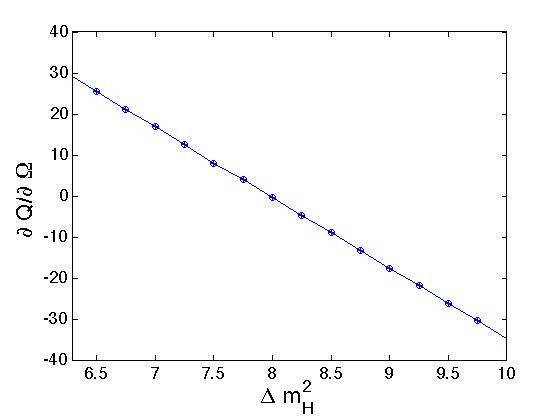} 
   \caption{Stability criterion ${\partial Q_s \over \partial \Omega_s} $ at $\Omega_s=m_W= {\sqrt 2 \over 2}$. We can see a change in sign occuring at $\Delta m_H^2 \approx 8$, signaling the transition from stability to instability.  } 
   \label{fig:stab}
\end{figure}

We calculate the derivative ${\partial Q_S \over \partial \Omega_S} >0$ at the point  $\Omega_S = m_W $ and the results are shown in Fig. \ref{fig:stab}. From this plot we can read the value of the critical value of the mass mismatch parameter, referred to as the stability point, which is $\Delta m_H^2 \approx 8$.

\begin{figure}
\centering
\begin{tabular}{cc}
\includegraphics[width=3.5in]{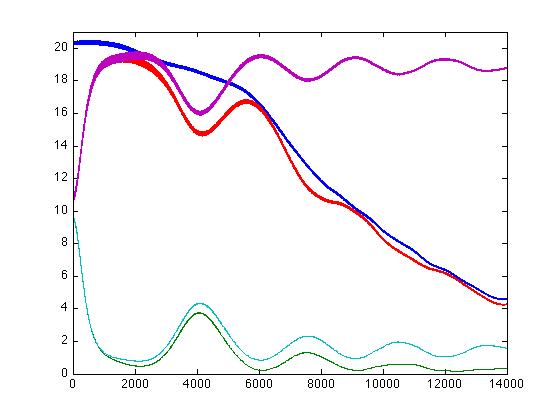}&
\includegraphics[width=3.5in]{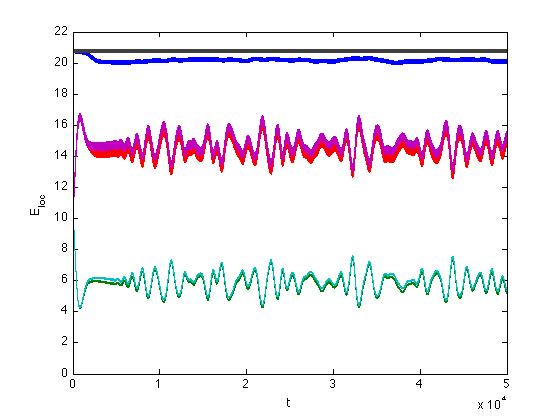} 
\end{tabular}
   \caption{Evolution curves for $\lambda=1-1.2\epsilon^2=0.997$ (left) and $\lambda = 1-0.8\epsilon^2=0.998$ (right) for $\epsilon=0.05$,  the two sides of the $\lambda=1- \epsilon^2$ line. Initial profiles chosen to be equal to the profile at $\lambda=1$. We plot the total localized energy (blue), the localized Higgs energy (red), the total Higgs energy (purple), the localized W energy (green) and the total W energy (cyan). We see either a pumping of the W field (left) or a small shed of energy followed by a stable oscillon (right). }
   \label{fig:lightHiggsthres}
\end{figure}

Examining the two sides of the $2:1$ ratio separately and starting from the light Higgs side, we have stability for $\Delta m_H^2 > -4m_W^2$ and no oscillon (meaning also instability) for $\Delta m_H^2< -4m_W^2$. This translates into a stability threshold at $\lambda = 1-\epsilon^2$, which for $\epsilon=0.05$ is placed at $\lambda_{threshold} = 0.9975$. This is numerically verified by taking two values very close to this point in Fig.\ref {fig:lightHiggsthres}. The initial conditions are taken to be equal to the oscillon profile for $\lambda=1$. In the stable region, we see the oscillon shedding some energy initially and settling to a stable configuration, while in the unstable region, we see a very quick pumping of the Higgs field, which then disperses.

\begin{figure}
\centering
\begin{tabular}{cc}
\includegraphics[width=3.5in]{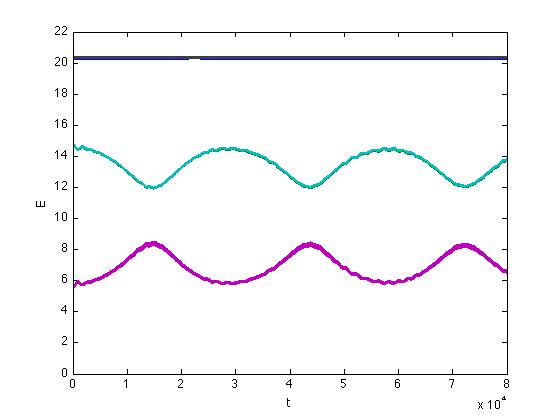}&
\includegraphics[width=3.5in]{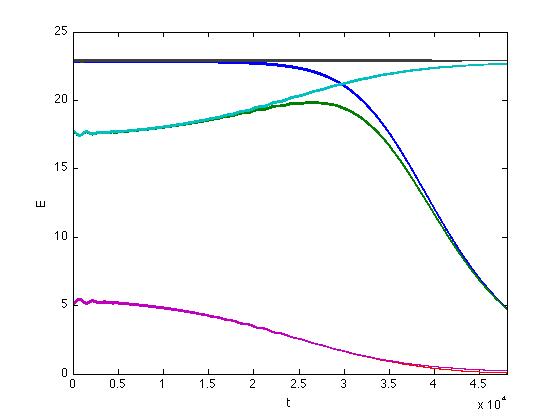} 
\end{tabular}
   \caption{Evolution curves for $\lambda=1+3\epsilon^2\approx 1.0019 $ (left) and $\lambda= 1+4.5\epsilon^2 \approx1.0028 $ (right) for $\epsilon=0.025$, the two sides of the $\lambda=1+4\epsilon^2$ line. Color coding as before. We can see either a stable oscillon (left) with a periodic energy exchange among its constituents, or a very efficient pumping of the Higgs field followed by a diffusive W decay (right).  }
   \label{fig:heavyHiggsthres}
\end{figure}

Doing the same for a heavy Higgs we find the stability threshold at $\lambda \approx 1+ 4\epsilon^2$, based on Fig. \ref{fig:stab}, which for $\epsilon=0.025$ is placed at $\lambda_{threshold} = 1.0025$. This is numerically seen by taking two values very close to this point in Fig.\ref {fig:heavyHiggsthres}. The profiles are taken to be the ones derived from the envelope equations including the $\Delta m_H^2$ term. For the stable case we see that the total energy curve and the localized energy curve for each field and the oscillon are indistinguishable from each other. The localized oscillon energy is constant, while the two fields periodically exchange energy. For the unstable case, the Higgs transfers all its energy into the W, which then disperses.

\section{Conclusion}

We provided an extensive investigation of oscillons in the reduced Spherical Ansatz in the SU(2) gauged Higgs model and explained their behavior for different values of the Higgs mass and the expansion parameter. 

We can divide the $\epsilon -\lambda$ plane in two ways, the criteria being either the stability of oscillons and oscillon-like configurations or the formal existence of oscillon solutions in the small amplitude expansion.

Initially we divide the regions of the mass mismatch based on the formal existence of solutions within the $\epsilon$ expansion framework.

\begin{itemize}
\item There is a single small amplitude expansion in the case $\lambda = 1+O(\epsilon^2)$.
\item The region $\lambda = 1+O(\epsilon)$ does not contain a small-amplitude two-timing expansion.
\item Finally, there are two scalings, leading to two sets of equations in the case $m_H-2m_W = O(1)$. The symmetric scaling leads to fields with amplitudes large enough to probe both minima of the potential. The asymmetric scaling leads to an oscillon consisting predominantly by the Higgs. All these formal oscillon solutions are unstable when evolved under the full wave equations.

\end{itemize}

The region of interest when searching for stable oscillons is the region near the stable point, where furthermore there is a well-defined small amplitude expansion, resulting to a set of coupled Nonlinear Schroedinger type equations.

The $(\epsilon , \lambda)$ plane can be divided in regions depending on the behavior of the oscillon (or oscillon-like) solutions. 

\begin{figure}[] 
   \includegraphics[width=6in]{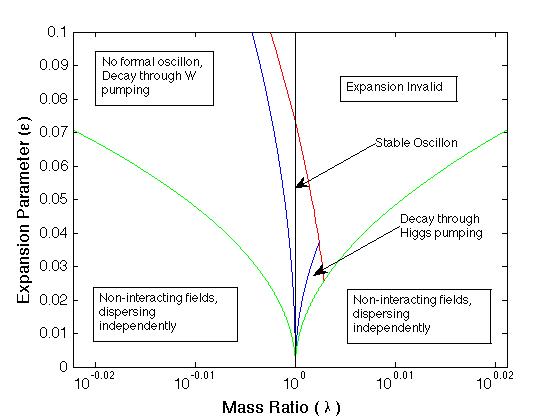} 
   \caption{The different regions of the $\lambda$-$\epsilon$ plane. The red curve is the $\epsilon_{max}$ line, above which the small amplitude expansion breaks down. The two outer green curves $\lambda = 1\pm 10 \epsilon^2$ define approximately the transition from the interacting to the dispersing region. The two inner blue curves   $\lambda = 1 - \epsilon^2$ and $\lambda = 1+4 \epsilon^2$ define the boundaries of the oscillon stability region. } 
   \label{fig:plane}
\end{figure}

\begin{itemize}
\item The large $\epsilon$ region defined as $1\over max\{\beta(\rho)\}$, where oscillons are not stable, due to the violation of the small amplitude nature of the expansion.

\item The region where the oscillon fields are not interacting, due to the fact that the relevant Fourier modes are exponentially suppressed by the form of the oscillon (or any other exponentially localized lump of size $O({1\over \epsilon})$. All tested oscillon-like configurations in this region are found to be unstable, which each component dispersing independently.

\item The region, where there is interaction and possibly a stable phase-locked oscillon configuration.
This in turn is divided into a stable and an unstable part. 

The characteristic dip in the lifetime graph of the oscillons, corresponds to a change in the behavior of the oscillon at $\lambda \approx 1+10\epsilon^2 $. This is the point where the $\lambda = 1+O(\epsilon^2)$ approximation starts to lose its validity.

An overview of the different regions is found in Fig. \ref{fig:plane}
\end{itemize}

We are currently applying our insight of this system into simpler models of composite oscillons, where the resulting equations have a simpler structure, leaving more room for exact Feynman graph calculations and a more rigorous stability analysis. 

\acknowledgements {We would especially like to thank Alan Guth for sharing his physical insight, Ruben Rosales for help with the stability analysis and Mark Hertzberg for suggestions about the presentation of the results. We also thank Mustafa Amin, Edward Farhi and David Shirokoff for helpful discussions. This work is supported by the U.S. Department of Energy (DoE) under Contract No. DE-FG02-05ER41360.}

\appendix

\section{Study of oscillons for $m_H-2m_W=O(1)$}

\subsection {Symmetric scaling}

In the case when the mass mismatch is of order one, one uses the appropriate scaling analysis leading to the rather more complicated equations described in the previous section. 

\begin{figure}
\centering
\begin{tabular}{cc}
\includegraphics[width=3.5in]{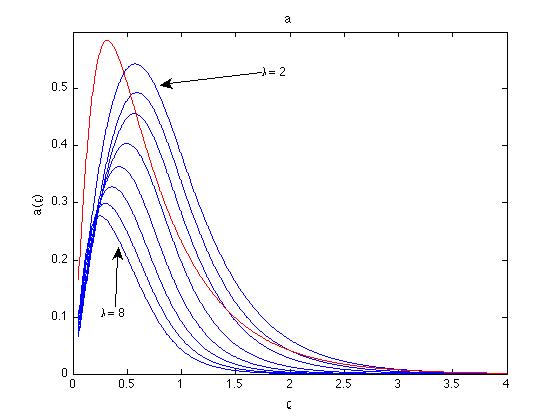}&
\includegraphics[width=3.5in]{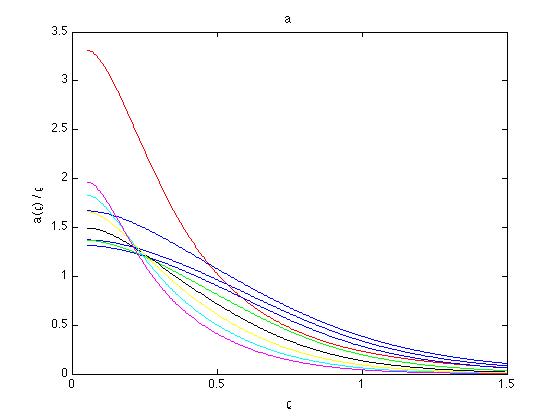} 
\end{tabular}
   \caption{Profiles of  $a(\rho)$ (left) and $a(\rho) / \rho$ (right) for different values of the mass ratio, $1.5\le \lambda \le 8$. Color-coding goes as follows: 1.5 red, 2, 2.5, 3 blue, 4 green, 5 black 6 yellow 7 cyan 8 magenta} 
   \label{fig:a_lamdaO1}
\end{figure}

\begin{figure}
\centering
\begin{tabular}{cc}
\includegraphics[width=3.5in]{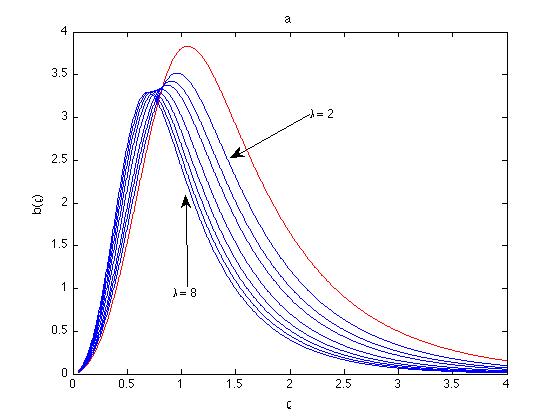}&
\includegraphics[width=3.5in]{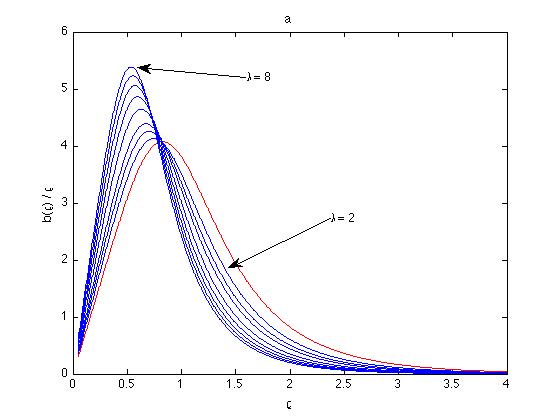} 
\end{tabular}
   \caption{Profiles of $b(\rho)$ (left) and $b(\rho) / \rho$ (right)  for different values of the mass ratio, $1.5\le \lambda \le 8$ } 
   \label{fig:b_lamdaO1}
\end{figure}

Starting with the symmetric scaling, we only numerically computed the node-less (fundamental) solution in the region of $\lambda$ varying ftom $1.5$ to $8$. With the value of $m_W=\sqrt 2 /2$, this leads to $ 0.3 \lessapprox m_H-2m_W \lessapprox 2.6$. The resulting profiles are plotted in Fig. \ref{fig:a_lamdaO1} and Fig. \ref{fig:b_lamdaO1}.  We see that in both ways of plotting $b(\rho)$, the curves are smoothly changed as $\lambda$ varies. On the other hand, in the case of $a(\rho)$ such a smooth change is not observed, especially when one plots $a(\rho) / \rho$.

The amplitudes for all values of the Higgs mass in this region exceed unity and are not rescaled by a small amplitude parameter. This leads to the fields probing both minima of the Higgs potential, leading to the quick decay of the oscillon.

\subsection {Asymmetric scaling}

\begin{figure}
\centering
\begin{tabular}{cc}
\includegraphics[width=3.5in]{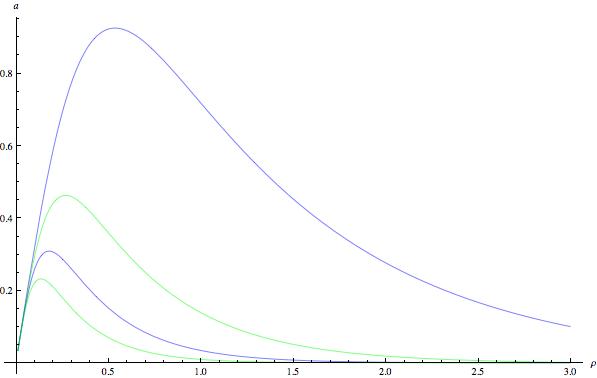}&
\includegraphics[width=3.5in]{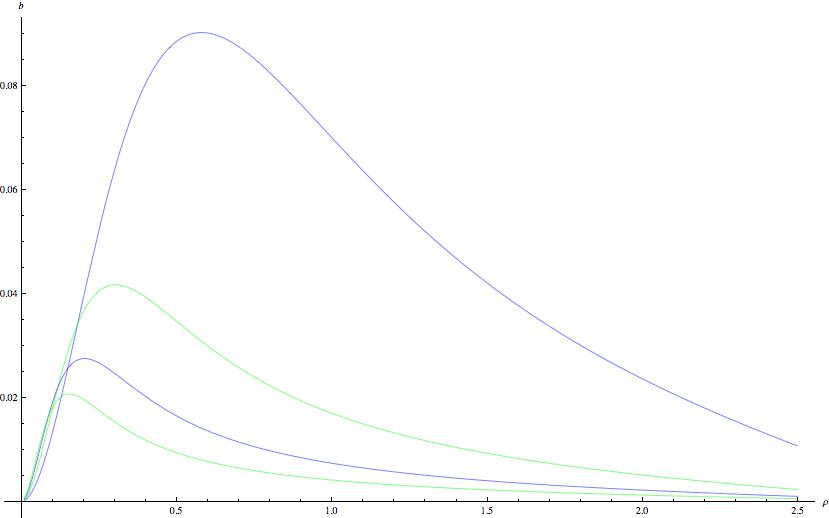} \\
\end{tabular}
   \caption{Profiles of $a(\rho)$ (left) and $b(\rho)$ (right) for $m_H=1,2,3,4$.} 
   \label{fig:unit_mismatch_2}
\end{figure}

For the asymmetric scaling, we again calculated only the node-less solution for $m_H = 1,2,3,4$, shown in Fig. \ref{fig:unit_mismatch_2}. As the mass increases both field amplitudes get smaller. The amplitude in the Higgs field is larger, and when taking into account that the Higgs amplitude is not $\epsilon$-suppressed, this is almost a single-field oscillon.

In the $m_H-2m_W=O(1)$ region the asymmetric scaling $\delta_\xi = O(1)$ and $\delta_\eta=\epsilon$ is only the first in an infinite ladder of scalings, namely  $\delta_\xi = O(1)$ and $\delta_\eta=\epsilon^n$ where $n\ge1$. The resulting equations from all these scalings are the same, which means the resulting physical oscillon is the same for the Higgs field and the same rescaled by a different power of $\epsilon$ for the W field. All these scalings point to an oscillon composed predominately by the Higgs field. We will therefore examine the limiting case of this ladder of scalings, $n\to \infty$. Simply put, we will set the W field to zero and consider the Higgs in its double-well potential. The only difference between this oscillon and other single scalar field oscillons in double well potentials found in the literature is that in our case it has space-dependent coefficients, resulting from the dimensional reduction due to the use of the Reduced Spherical Ansatz.

The equation of motion of the Higgs field in the absence of the W field is 
\begin{eqnarray}\partial^\mu \partial_\mu \xi +2\lambda v^2 r \left [ {\xi \over r} -{3\over 2} \left ({\xi \over r}\right )^2 +{1\over 2} \left ({\xi \over r}\right )^3 \right ] =0\end{eqnarray}
This can be derived from the equation of motion for the physical Higgs field
\begin{eqnarray}\left [ D^\mu r^2 D_\mu  +{2\lambda \over g^2} r^2 \left ( |\phi|^2-{g^2v^2\over 2} \right ) \right ] \phi= ( \phi^*-\phi)  \end{eqnarray}
by the substitution $\phi = {gv\over \sqrt 2} \left ( 1-{\xi \over r }\right )$

Following the $\epsilon$-expansion the $\xi = Re \{ A(\rho,\tau)e^{-im_Ht} \}$ where $A(\rho,\tau)$ is defined by
\begin{eqnarray}-2im_H {\partial A\over \partial \tau }= {\partial ^2 A \over \partial \rho ^2} + {14 m_H ^4\over 8 \rho ^2} |A|^2A\end{eqnarray}
By using separation of variables $A(\rho,\tau) = a(\rho) e^{i\Omega \tau}$ the amplitude $a(\rho)$ is calculated by 
\begin{eqnarray} {\partial ^2 A \over \partial \rho ^2} - 2m_H\Omega A+ {14 m_H ^4\over 8 \rho ^2} |A|^2A = 0\end{eqnarray}
A similar stability analysis as before, leads to the oscillons being stable for 
\beq
{\partial Q \over \partial \Omega} >0
\eeq
 where 
\beq
Q=\int _0 ^\infty d\rho |A|^2
\eeq
Calculations for various values of $m_H$ and $\Omega$ show a globally unstable oscillon.

The physical Higgs field depends on ${\xi \over r}$. By setting 
\beq 
\tilde \xi = {\xi \over r}
\eeq
the equation of motion becomes
\beqn {\partial^2 \tilde \xi \over \partial t^2} - {\partial^2 \tilde \xi \over \partial r^2} - {2\over r} {\partial \tilde \xi \over \partial r}+2\lambda v^2  \left [ \tilde \xi -{3\over 2} \tilde \xi^2 +{1\over 2} \tilde \xi^3 \right ] =0
\eeqn
This is the equation of motion of a radial oscillon in 3 spatial dimensions with constant coefficients. It is well documented that such an oscillon will be unstable. Thus in the reduced spherical Ansatz the Higgs field forms an unstable 3 dimensional oscillon, that can be stabilized near the $2:1$ mass ratio through interaction with the W field. The interesting question arises, under what conditions an unstable oscillon can be stabilized through interaction with a second field. The second field could even have a different dimensionality, meaning a radial laplacian with a different coefficient of the first derivative term. There are no general results to our knowledge regarding the interaction of two fields with (effectively) different dimensionality within an oscillon. Investigations in this direction will be carried out in the future.

\section{Space and Time Scaling}

Even though we used general scaling for the field amplitudes, the scaling for the space and time were chosen to be $\epsilon$ and $\epsilon^2$ respectively. One of these can be chosen arbitrarily, since $\epsilon$ does not have any meaning other than being a small number. Let us consider a single scalar field with an equation of motion of the form 
\beq
{\partial^2 u \over \partial t^2} - {\partial^2 u \over \partial r^2} +m^2 u +\{nonlinear~ terms\}=0
\eeq
If one chooses $\rho = \epsilon r$, the e.o.m. becomes 
\beq
{\partial^2 u \over \partial t^2} -\epsilon^2 {\partial^2 u \over \partial \rho^2} +m^2 u +\{nonlinear~ terms\}=0
\eeq
Keeping the scaling of the slow time general, say $\tau = \epsilon^\kappa t$, we get
\beq
{\partial^2 u \over \partial t^2} + 2\epsilon ^\kappa {\partial u \over \partial \tau}+{\partial^2 u \over \partial \tau^2}  -e^2 {\partial^2 u \over \partial \rho^2} +m^2 u +...=0
\eeq
If we want the slow time and the space dependence to enter at the same order, we must choose $\kappa=2$, hence $\tau = \epsilon^2 t$. This conclusion is general and does not change for interacting fields and any type of nonlinearity.

With this final note we have exhausted all possible oscillons in the reduced spherical Ansatz of the $SU(2)$ gauged Higgs model.


\begin{thebibliography}{99}

\bibitem {tail} H.~Segur and M.~D.~Kruskal, 
``Nonexistence of small- amplitude breather solutions in $\phi^4$ theory,''
 Phys. Rev. Lett. {\bf 58}, 747 - 750 (1987)

\bibitem{radnum} P.~Salmi and M.~Hindmarsh,
  ``Radiation and Relaxation of Oscillons,''
  Phys.\ Rev.\ D {\bf 85}, 085033 (2012)
  [arXiv:1201.1934 [hep-th]]
  
  \bibitem {rad1} G.~Fodor, P.~Forgacs, Z.~Horvath and M.~Mezei,
  ``Radiation of scalar oscillons in 2 and 3 dimensions,''
  Phys.\ Lett.\ B {\bf 674}, 319 (2009)
  [arXiv:0903.0953 [hep-th]]

\bibitem{rad2} G.~Fodor, P.~Forgacs, Z.~Horvath and M.~Mezei,
  ``Radiation of scalar oscillons in 2 and 3 dimensions,''
  Phys.\ Lett.\ B {\bf 674}, 319 (2009)
  [arXiv:0903.0953 [hep-th]]

\bibitem {exp} N.~Graham and N.~Stamatopoulos,
  ``Unnatural Oscillon Lifetimes in an Expanding Background,''
  Phys.\ Lett.\ B {\bf 639}, 541 (2006)
  [hep-th/0604134]

\bibitem{mark} M.~P.~Hertzberg,
  ``Quantum Radiation of Oscillons,''
  Phys.\ Rev.\ D {\bf 82}, 045022 (2010)
  [arXiv:1003.3459 [hep-th]]

\bibitem{theory1} M.~Gleiser and D.~Sicilia,
  ``A General Theory of Oscillon Dynamics,''
  Phys.\ Rev.\ D {\bf 80}, 125037 (2009)
  [arXiv:0910.5922 [hep-th]]

\bibitem{theory2} M.~Gleiser and D.~Sicilia,
  ``Analytical Characterization of Oscillon Energy and Lifetime,''
  Phys.\ Rev.\ Lett.\  {\bf 101}, 011602 (2008)
  [arXiv:0804.0791 [hep-th]]

\bibitem{theory3} G.~Fodor, P.~Forgacs, Z.~Horvath and A.~Lukacs,
  ``Small amplitude quasi-breathers and oscillons,''
  Phys.\ Rev.\ D {\bf 78}, 025003 (2008)
  [arXiv:0802.3525 [hep-th]]



\bibitem{infl1} M.~A.~Amin, R.~Easther, H.~Finkel, R.~Flauger and M.~P.~Hertzberg,
  ``Oscillons After Inflation,''
  Phys.\ Rev.\ Lett.\  {\bf 108}, 241302 (2012)
  [arXiv:1106.3335 [astro-ph.CO]]
  
  \bibitem{infl2} M.~A.~Amin, R.~Easther and H.~Finkel,
  ``Inflaton Fragmentation and Oscillon Formation in Three Dimensions,''
  JCAP {\bf 1012}, 001 (2010)
  [arXiv:1009.2505 [astro-ph.CO]]
  
\bibitem{infl3} M.~A.~Amin,
  ``Inflaton fragmentation: Emergence of pseudo-stable inflaton lumps (oscillons) after inflation,''
  arXiv:1006.3075 [astro-ph.CO]

\bibitem{bubble} E.~J.~Copeland, M.~Gleiser and H.~-R.~Muller,
  ``Oscillons: Resonant configurations during bubble collapse,''
  Phys.\ Rev.\ D {\bf 52}, 1920 (1995)
  [hep-ph/9503217]
  
\bibitem{emerge} E.~Farhi, N.~Graham, A.~H.~Guth, N.~Iqbal, R.~R.~Rosales and N.~Stamatopoulos,
  ``Emergence of Oscillons in an Expanding Background,''
  Phys.\ Rev.\ D {\bf 77}, 085019 (2008)
  [arXiv:0712.3034 [hep-th]]

\bibitem{bath} M.~Gleiser and R.~M.~Haas,
  ``Oscillons in a hot heat bath,''
  Phys.\ Rev.\ D {\bf 54}, 1626 (1996)
  [hep-ph/9602282]

\bibitem {composite}  M.~Gleiser, N.~Graham and N.~Stamatopoulos,
  ``Generation of Coherent Structures After Cosmic Inflation,''
  Phys.\ Rev.\ D {\bf 83}, 096010 (2011)
  [arXiv:1103.1911 [hep-th]]

\bibitem{num} E.~Farhi, N.~Graham, V.~Khemani, R.~Markov and R.~Rosales,
  ``An Oscillon in the SU(2) gauged Higgs model,''
  Phys.\ Rev.\ D {\bf 72}, 101701 (2005)
  [hep-th/0505273].
  
  \bibitem{su2} E.~Farhi, N.~Graham, A.~H.~Guth, R.~R.~Rosales and R.~Stowell, 
  ``Constructing Oscillons in a Reduced Spherical Ansatz for the $SU(2)$ Gauged Higgs Model,''
  to be published
  
\bibitem{noah1}N.~Graham,
  ``An Electroweak oscillon,''
  Phys.\ Rev.\ Lett.\  {\bf 98}, 101801 (2007)
  [Erratum-ibid.\  {\bf 98}, 189904 (2007)]
  [hep-th/0610267]
  
\bibitem{noah2} N.~Graham ,
  ``Numerical Simulation of an Electroweak Oscillon,''
  Phys.\ Rev.\ D {\bf 76}, 085017 (2007)
  [arXiv:0706.4125 [hep-th]]

\bibitem{su2cosmo} M.~Gleiser, N.~Graham and N.~Stamatopoulos,
  ``Long-Lived Time-Dependent Remnants During Cosmological Symmetry Breaking: From Inflation to the Electroweak Scale,''
  Phys.\ Rev.\ D {\bf 82}, 043517 (2010)
  [arXiv:1004.4658 [astro-ph.CO]]



\bibitem{ansatz} B.~Ratra and L.~G.~Yaffe,
  ``Spherically Symmetric Classical Solutions In Su(2) Gauge Theory With A Higgs Field,''
  Phys.\ Lett.\ B {\bf 205}, 57 (1988)

\bibitem{ginv}E.~Farhi, K.~Rajagopal and R.~L.~Singleton, Jr,
  ``Gauge invariant variables for spontaneously broken SU(2) gauge theory in the spherical ansatz,''
  Phys.\ Rev.\ D {\bf 52}, 2394 (1995)
  [hep-ph/9503268]

\bibitem{stab} A.~V.~Buryak, P.~Di~Trapani, D.~V.~Skryabin and S.~Trillo, 
``Optical solitons due to quadratic nonlinearities: from basic physics to futuristic applications,'' Physics Reports {\bf 370}, 63-235 (2002) 

\bibitem{stab2} D.~E.~Pelinovsky, A.~V.~Buryak and Yu.~S.~Kivshar,
 ``Instability of Solitons Governed by Quadratic Nonlinearities,'' Phys.Rev.Lett. {\bf 75}, 591-595 (1995)
 
\bibitem{VK} N.~G.~Vakhitov and A.~A.~Kolokolov, 
``Stationary solutions of the wave equation in a medium with nonlinearity saturation,'' Radiophys. Quantum Electron. {\bf 16}, 783 (1975)

\bibitem{VK2}A.~A.~Kolokolov,
 ``Stability of stationary solutions of nonlinear wave equations,'' Radiophys. Quantum Electron.  {\bf 17}, 1016 (1976)




























\end{thebibliography}
\end{document}